\documentclass[11pt,a4paper]{article}
\usepackage[english]{babel}
\usepackage{a4wide}
\usepackage[latin1]{inputenc}
\usepackage{amsmath}
\usepackage{amsfonts}
\usepackage{amssymb}
\usepackage{epsfig}

\usepackage{graphicx}
\usepackage{latexsym}
\usepackage{color}

\title{Launchers and Targets in Social Networks}
\author{Pedro Martins \\ Polytechnic Institute of Coimbra, Coimbra Business School - ISCAC, Portugal, and \\ Centro de Matemática, Aplicações Fundamentais e Investigação Operacional \\ (CMAFcIO), Universidade de Lisboa, 1749-016 Lisboa, Portugal \\
e-mail: \texttt{pmartins@iscac.pt}
\and Filipa Alarcão Martins \\ Digital Marketing Analyst, Portugal \\
e-mail: \texttt{filalarcaomartins@hotmail.com}}

\begin{document}

\maketitle \setcounter{page}{1}

\begin{abstract}
Influence propagation in social networks is a subject of growing interest. A relevant issue in those networks involves the identification of key influencers. These players have an important role on viral marketing strategies and message propagation, including political propaganda and fake news. In effect, an important way to fight malicious usage on social networks is to understand their properties, their structure and the way messages propagate.

This paper proposes two new indices for analysing message propagation in social networks, based on the network topological nature and the power of the message. The first index involves the strength of each node as a launcher of the message, dividing the nodes into launchers and non-launchers. The second index addresses the potential of each member as a receiver (target) of the message, dividing the nodes into targets and non-targets. Launcher individuals should indicate strong influencers and target individuals should identify the best target consumers. These indices can assist other known metrics when used to select efficient influencers in a social network. For instance, instead of choosing a strong and probably expensive member according to its degree in the network (number of followers), we may previously select those belonging to the launchers group and look for the lowest degree members, which are probably cheaper but still guarantying almost the same influence effectiveness as the largest degree members.

On a different direction, using the second index, the strong target members should characterize relevant consumers of information in the network, which may include fake news' regular collectors.

We discuss these indices using small-world randomly generated graphs and a number of real-world social networks available in known datasets repositories.



\end{abstract}

\section{Introduction} \label{Sec1:Intro}

Social networks have long existed in society, but the fast growth of the web made these networks emerge at an unimaginable scale. These networks represent linkage among people and besides their relevancy in communication and in society, they also provide influence exertion, information dissemination (true or false) and, in some cases, gossip spread.

Message or information propagation follows on a cascade setting in an online social network. As an example, suppose that John writes on a friend's Facebook wall about a party he has been last night. This post is communicated to his friends who may typically comment, getting the message across their friends, and so on. This way, John's initial message propagates transitively throughout the network. In another example, suppose that Mary posts on Twitter about a new mobile phone she bought. Some of her followers on Twitter reply to her tweet while others retweet it, producing again a cascade of information propagation. In these examples, the messages start on a single person as is usual in many posts and comments on social networks. Our study will focus on these cases, where a message is launched by a single user. In these cases, an interesting point to observe is how far the message can go throughout the network when sent by each of the users; and who are the most targeted users in the network. These are the aspects that we propose discussing in the present work. Along the text, we will use the word "message" in a broad sense, representing a content launched by a member into the network.

This paper considers a social network on a simple and oriented graph (or network) $G=(N,A)$, with $N=\{1, \ldots, n\}$ the set of nodes or individuals or members (homogeneous) and $A \subseteq \{(i,j) \in N \times N: i \neq j \}$ the set of arcs, representing the existing links among individuals. We denote by $\delta^-(j)$ and $\delta^+(j)$ the set of predecessors and the set of successor of $j$ in $G$, respectively. We also denote by $g^-(i)$ and $g^+(i)$ the in-degree and out-degree of node $i$, respectively, thus, $g^-(i) = |\delta^-(i)|$ and $g^+(i) = |\delta^+(i)|$. Arc $(i,j) \in A$ indicates that node $i$ influences node $j$. In the context of a social network, this means that individual $i$ is followed by individual $j$ (note that the orientation of the arc can seem contradictory, however, we are assuming that an individual influences its followers, as is usual). A node is said to be a \emph{seed} if it is the starting point of a message launched into the network, meaning that this node/individual is the launcher of a message. The \emph{strength of influence} that node $i$ exerts on $j$ is characterized by $d(i,j)$, for all $(i,j) \in A$; and the \emph{hurdle} of each node $j \in N$ to adopt the message is denoted by $h(j)$. Thus, following other authors notation, node $j$ is said to be \emph{active} (or \emph{covered}) if it is a seed node or if $\sum_{i \in S \cap \delta^-(j)} d(i,j) \geq h(j)$, for $S \subseteq N$ the current subset of active nodes. This expression is denoted by \emph{activating condition}. So, in an online marketing setting for promoting a product, a node/individual is active if it has adopted or promoted the product, otherwise it is inactive. In this study we assume that an active member is an adopter. A more detailed discussion about the roles of an influencer node is conducted in \cite{CheLakCas2013}. In addition, if the original graph is undirected, then each edge $\{i,j\}$ should be substituted by the two arcs $(i,j)$ and $(j,i)$.

The activating condition previously described is based on the Linear Threshold (LT) model proposed in \cite{KemKleTar2003}, where a person will adopt a product/message if the influence received from its neighbors has reached a certain threshold (the hurdle value of the node). This condition follows similar expressions used on influence propagation and influence maximization problems in the literature, namely in \cite{Che2009}, and in very recent publications in \cite{FisKahLeiMonRut2018}, \cite{RagZha2019} and \cite{GunRagZha2020}. All these works address propagation originated on a set of members and not on a single individual as in our case. Finding such a set of the smallest size that would lead the entire network to repass the message or adopt the product is known as the Target Set Selection (TSS) problem, discussed in \cite{Che2009}. In the TSS, 100\% adoption is required, the hurdle $h(i)$ of a node $i \in N$ is a value between 1 and its degree in $G$ and a node $i$ becomes active if it has at least $h(i)$ active neighbors, that is, equal influence is assumed. Later on, \cite{RagZha2019} considered a weighted version of the TSS, denoted by WTSS problem. In the WTSS, equal influence from neighbors and 100\% adoption is still present, but the hurdle of a node is a value (a weight) that characterizes the amount of effort required by an individual (node) to be convinced to adopt the product. Equal influence and 100\% adoption are no longer required on the Least-Cost Influence Problem (LCIP) discussed in \cite{GunRagZha2020}. In the LCIP, the activating condition includes a tailored (i.e., partial) incentive (monetary in most cases) on each node, exerting influence on that node together with the usual influence employed by its predecessors to promote adoption. The LCIP seeks minimizing the sum of all tailored incentives provided to individuals in a social network while ensuring that a given fraction of the network is influenced. This problem was also addressed in \cite{FisKahLeiMonRut2018}, considering a nonlinear influence structure, besides the usual settings that characterize the LCIP.

An extensive survey on influence propagation in social networks is described in \cite{PenZhoCaoYuNiuJia2018}. It encompasses various types of social networks, their properties, social influence, state-of-the-art evaluation metrics and models, and an overview of known methods for influence maximization.

The interpretation of the strength of influence $d(i,j)$ that individual $i$ exerts on $j$ is not a theme of consensus, as observed in most works previously reported on influence propagation and influence maximization. In fact, it is not easy to set a function for characterizing the influence parameters $d(i,j)$, neither the hurdle of a node as in most cases it is individual dependent, characterized by the message/product and it may also vary along the time. In our case, we set these parameters as functions of the nodes' degrees, and thus defining the activating condition as a linear function of nodes' degrees. This way, we do not have to assess influence strength and nodes' hurdles for each particular instance. Instead, the activating conditions are automatically build, translating the inherent topological nature of the network for characterizing influence propagation, with an additional single parameter to assess the viral power of the message/product involved. Thus, in the present work we consider that influence strength of an individual can be related with the number of other individuals that he/she directly exerts influence on. In this case, we define this influence as its out-degree if working on an oriented graph, or degree if the graph is undirected. The same way, the hurdle can be associated with the influence strength of the individual (its out degree) and with the influence/viral power of the message (parameter $\alpha$, denoted by \emph{hurdle coefficient}). It acts as a threshold for this individual to become active, that is, to adopt a product or to repass a message. Thus, we define:
\begin{itemize}
  \item $d(i,j) = g^+(i)$, for all $(i,j) \in A$, that is, $d(i,j)$ is the out-degree of node $i$, indicating that the strength of influence that $i$ exerts on $j$ is defined by the number of individuals that $i$ can influence, that is, the number of followers of node $i$; and
  \item $h(j) = \alpha \cdot g^+(j)$, for all $j \in N$, where $\alpha$ is the hurdle coefficient, used to leverage the difficulty to activate node $j$, being related with the viral power of the message sent by the seed; and $g^+(j)$ is, again, node´s $j$ own strength.
\end{itemize}

Hence, for any node $j \in N$ and $S \subseteq N$ the current subset of active nodes, the activating condition becomes:
\begin{equation}\label{eqn:1}
     \sum_{i \in S \cap \delta^-(j)} g^+(i) \geq \alpha \cdot g^+(j)
\end{equation}

According to the previous definition of the influence that a node $j \in N$ receives from their direct neighbors (characterized by parameters $d(i,j)$, for all $(i,j) \in A$), equal influence from predecessor neighbors is not present, except if $G$ is regular, which is quite unlikely in a social network. Equal influence from predecessors was considered on the TSS and WTSS problems previously mentioned, being unavoidable when privacy concerns are present in the network. In addition, the hurdle is also node dependent as in the WTSS and LCIP problems, indicating that different nodes require different levels of effort to become active. In this paper, we assume that this hurdle is proportional to the node's out-degree (its number of followers), as mentioned above. The relevancy of the number of followers is also stressed in the literature (see, e.g., \cite{Huaetal2012,BakHofMasWat2011}) and widely used on a number of public social activities, namely to select marketing influencers, or people to TV shows, or other public exhibitions, which motivates our option. This way, all the influential process is based on the topological nature of the graph, using just the degree information of the nodes in the entire activating condition. This excludes other external incident features, like monetary incentives, user specific characterizations of the influence among individuals or specific characterizations of the hurdle.

This paper proposes two new indices for analysing message propagation in social networks. These indices are based on two new concepts:

\begin{itemize}
  \item the \emph{Individual Launching Power} (ILP) of node $i \in N$, denoted by $ilp'(i)$, representing the number of activated nodes in $G$ when $i$ is the launcher of a message; and
  \item the \emph{Individual Target Potential} (ITP) of node $i \in N$, denoted by $itp'(i)$, representing the number of times that node $i$ is activated when each of the other nodes $j \in N \setminus \{i\}$ launch their own messages.
\end{itemize}

Thus, we define the following two indices, respectively:

\begin{itemize}
  \item the ILP index of $i$: $ilp(i) = \frac{ilp'(i)}{(n-1)}$, for all $i \in N$; and
  \item the ITP index of $i$: $itp(i) = \frac{itp'(i)}{(n-1)}$, for all $i \in N$.
\end{itemize}

Index $ilp(i)$ represents the potential strength of individual $i$ as a launcher of a message, while index $itp(i)$ is the potential of individual $i$ as a receiver (target) of a message sent from the nodes of $G$. Launcher individuals should correspond to strong influencers and target individuals should identify the best target consumers or the most prominent message collectors. These indices are assuming that all nodes are receptive for the message/product and all nodes have equal chance to be the seeds of a message/product.

There are three relevant issues involved on message propagation: i) the launcher strength, ii) the message power, and iii) the network topological structure. The launcher strength can be set by the ILP index, the message influence/viral power is characterized by the hurdle coefficient $\alpha$ and the network structure is the particular topology of graphs $G$. Note that the relevancy of classifying the viral power of the message was also stressed in \cite{BakHofMasWat2011} on the cascade size of influence propagation.

\begin{figure}[h]
\begin{center}
\begin{tabular}{ccccc}
graph $G1$ & & $\alpha = 1.5$ & & $\alpha = 2.0$ \\
\includegraphics[scale=0.50]{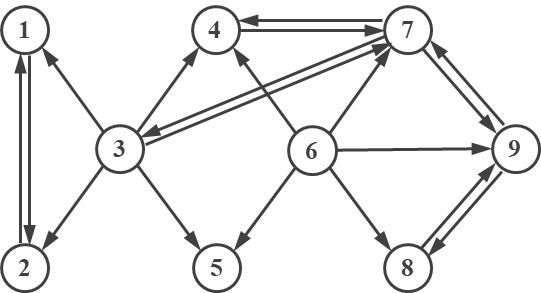} & \mbox{\hspace*{0,15cm}} & \includegraphics[scale=0.50]{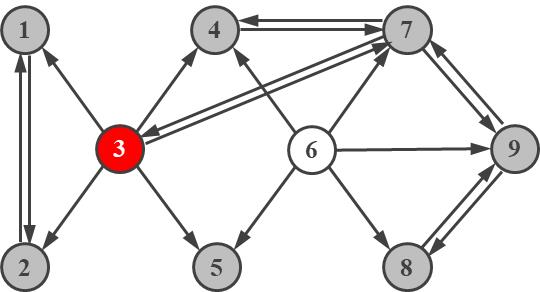} & \mbox{\hspace*{0,15cm}} & \includegraphics[scale=0.50]{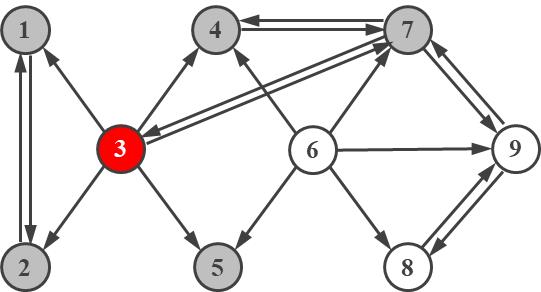} \\
(a) & & (b) & & (c) \\
\end{tabular}
\caption{\label{fig1} Oriented graph $G1$ in (a) and the active nodes when node 3 is the seed, for $\alpha = 1.5$ and $\alpha = 2.0$, in (b) and (c), respectively.}
\end{center}
\end{figure}

To exemplify, consider the oriented graph $G1$ in Figure \ref{fig1}a, with 9 nodes and 19 arcs.

Table \ref{tab:t1} shows the ILP and ITP indices' values for all the nodes in $G1$, for $\alpha = 1.5, 2.0 \mbox{ and } 2.5$. It also includes the out-degree of each node.

\smallskip {\small
\begin{table}[h]
  \centering
 {\footnotesize
 \begin{tabular}{c|c|ccc|ccc}
  \hline
   &  & \multicolumn{3}{|c}{$ilp(i)$} & \multicolumn{3}{|c}{$itp(i)$} \\
  node $i$ & $g^+(i)$ & $\alpha = 1.5$ & $\alpha = 2.0$ & $\alpha = 2.5$ & $\alpha = 1.5$ & $\alpha = 2.0$ & $\alpha = 2.5$ \\
  \hline
  1 & 1 & 0.000 & 0.000 & 0.000 & 0.125 & 0.125 & 0.125 \\
  2 & 1 & 0.000 & 0.000 & 0.000 & 0.125 & 0.125 & 0.125 \\
  3 & 5 & 0.875 & 0.625 & 0.500 & 0.000 & 0.000 & 0.000 \\
  4 & 1 & 0.000 & 0.000 & 0.000 & 0.375 & 0.375 & 0.375 \\
  5 & 0 & 0.000 & 0.000 & 0.000 & 0.250 & 0.250 & 0.250 \\
  6 & 5 & 0.625 & 0.625 & 0.625 & 0.000 & 0.000 & 0.000 \\
  7 & 3 & 0.375 & 0.125 & 0.125 & 0.250 & 0.250 & 0.125 \\
  8 & 1 & 0.000 & 0.000 & 0.000 & 0.500 & 0.250 & 0.125 \\
  9 & 2 & 0.125 & 0.125 & 0.000 & 0.375 & 0.125 & 0.125 \\
  \hline
\end{tabular}}
  \caption{{\small ILP and ITP indices for graph $G1$, including the nodes' out-degree ($g^+(i)$)}} \label{tab:t1}
\end{table}}

In this example and considering a message launched in the network by node 3, if the hurdle coefficient is $\alpha = 1.5$, representing a rather viral message, then it can reach all nodes but the 6, covering 87.5\% of the entire network (excluding node 3), represented in Figure \ref{fig1}b. In this case, node 7 is made active just by influence of node 3. Then, in turn, node 7 can influence node 9, which will then influence node 8. Alternatively, if node 3 releases a less viral message, with hurdle coefficient 2.0, then it only reaches 62.5\% of the other nodes, shown in Figure \ref{fig1}c. In this case, node 7 is activated through the joint influence of 3 and 4; and nodes 8 and 9 are no longer influenceable. However, if the hurdle coefficient is $\alpha = 2.5$, the message sent by 3 will reach only nodes 1, 2, 4 and 5, covering just 0.5\% of the nodes in $N \setminus \{3\}$. Besides, note that although nodes 3 and 6 have the same out-degree ($g^+(3) = g^+(6) = 5$), their influence power in the entire graph is different due to the influential cascade each one generates. In effect, in this case, node 3 can go further than node 6.

On the other hand, if observing the $itp(i)$ indices, representing the targeting potential of each node $i \in N$, when the message is very viral, with $\alpha = 1.5$, node 8 can be reached by half of the other individuals. However, when the message is less viral, with $\alpha = 2.0$ or $\alpha = 2.5$, the easiest individual to reach is node 4, which can be targeted by 3 of the remaining individuals (37.5\%).

If graph $G1$ is undirected, by substituting each arc by an edge between the same nodes, the connectivity in the network may seem extended. However, considering again node 3 as the seed of a message, the ILP indices are $ilp(3)=1$ for $\alpha=1.5 \mbox{ and } 2.0$, and $ilp(3)=0.375$ when $\alpha=2.5$. While for $\alpha=1.5 \mbox{ and } 2.0$ the influence capability increases; it shrinks instead when the message becomes less viral ($\alpha=2.5$). The reason for the influence reduction in the last case is related with the hurdle increase in most nodes, due to the out-degree growth, making some nodes harder to cover when the message is less viral.

The various social networks available these days vary quite a lot concerning oriented and undirected graph topologies. For example, Facebook and LinkedIn should be represented by undirected graphs, as each connection requires both users to accept the link. On the other hand, Instagram and ResearchGate, for instance, should be described using directed graphs, as user A can decide to follow B, but B may not be interested in following A.

The contribution of the present paper involves the following three aspects:
\begin{itemize}
  \item We propose two new indices especially devoted for identifying strong influencers and strong targeting nodes in a social network, through the ILP and the ITP indices, respectively. These indices incorporate a hurdle criteria on each node to model its hardness to forward the message, while most link topological measures in the literature are entirely build upon the relevancy of neighbors, ignoring the nodes' ability to decide on a message propagation scheme.
  \item The construction of the new indices uses just the underlying topological nature of the graph and a hurdle coefficient ($\alpha$) to characterize the viral strength of the message/product.
  \item The ILP index results divide the nodes into launchers and non-launchers. The launchers class can be used to assist other metrics to find the best influencers, providing better decisions. The ITP results divide the nodes into targets and non-targets, allowing to identify strong consumers of information that may possibly magnetize malicious information in the network.
\end{itemize}

The paper is organized as follows: an algorithm for constructing the ILP and ITP indices is described in the next section; computational tests are conducted in Section \ref{Sec3:Compt}; and the paper ends with a conclusions section (Section \ref{Sec4:Conclusions}).

\section{Algorithm for calculating the ILP and ITP indices} \label{Sec2:Alg}

The construction of the two indices (ILP and ITP) is made in the same algorithm, on a given simple oriented graph $G=(N,A)$. It calculates the $ilp(i)$ index for each node $i \in N$ and updates the $itp(i)$ values along the calculation of the ILP indices. So, starting from a given node $i \in N$ and for a given input parameter value $\alpha$, the algorithm generates a sequential cascade of newly active nodes through influence propagation, using the activating condition set in (\ref{eqn:1}). Its pseudocode is depicted in Figure \ref{Alg1}. Before, we describe the sets and the variables involved.

\medskip
\begin{tabular}{ll}
$S$ & is the current subset of activated nodes ($S\subset N$) \\
$L$ & is the current list of nodes to analyse (still non activated nodes) \\
$\delta^-(i)$ & is the set of nodes converging to $i$ in $G$, for all $i \in N$ \\
$\delta^+(i)$ & is the set of nodes diverging from $i$ in $G$, for all $i \in N$ \\
$g^+(i)$ & is the out-degree of node $i$ in $G$, for all $i \in N$ (that is, $g^+(i) = |\delta^+(i)|$) \\
$\alpha$ & is the hurdle coefficient \\
\end{tabular}
\medskip

\begin{figure}[h]
\begin{center}
\begin{tabular}{|lll|}
\hline
  & \texttt{\textbf{Algorithm ILP-ITP($\alpha$)}} & \\
  & \texttt{\textbf{Input:} $G=(N,A)$ and $\alpha$} & \\
  & \texttt{\textbf{Output:} $ilp(i)$ and $itp(i)$ for all $i \in N$} & \\
1 & \textbf{for} all $i \in N$ \textbf{do} & \\
2 & \hspace*{0.5cm} $itp(i) \leftarrow 0$ & //\emph{initialize vector itp}// \\
3 & \textbf{end\_do} & \\
4 & \textbf{for} all $i \in N$ \textbf{do}  & \\
5 & \hspace*{0.5cm} $S \leftarrow \{i\}$, $L \leftarrow \{j \in N \setminus S : (i,j) \in A\}$ & //\emph{initialize the sets}// \\
6 & \hspace*{0.5cm} \textbf{while} ($L \neq \emptyset$) \textbf{do} & \\
7 & \hspace*{1.0cm} $v \leftarrow \arg \min_{j \in L} \{g^+(j)\}$ & //\emph{select the next candidate to activate}// \\
8 & \hspace*{1.0cm} \textbf{if} $\left( \sum_{j \in S \cap \delta^-(v)} g^+(j) \geq \alpha \cdot g^+(v) \right)$ \textbf{then} & //\emph{if it becomes active, then}// \\
9 & \hspace*{1.5cm} $S \leftarrow S \cup \{v\}$ & //\emph{put the newly activated node in $S$}// \\
10 & \hspace*{1.5cm} $L \leftarrow L \cup \{j \in N\setminus S: (v,j) \in A \}$ & //\emph{add to $L$ all inactive successors of $v$}// \\
11 & \hspace*{1.5cm} $itp(v) \leftarrow itp(v) + 1$ & //\emph{$i$ is also able to activate node $v$}// \\
12 & \hspace*{1.0cm} \textbf{end\_if} & \\
13 & \hspace*{1.0cm} $L \leftarrow L \setminus \{v\}$ & //\emph{remove node $v$ from $L$}// \\
14 & \hspace*{0.5cm} \textbf{end\_while} & \\
15 & \hspace*{0.5cm} $ilp(i) \leftarrow |S|-1$ & //\emph{$|S|$ is the number of nodes that node $i$ can} \\
 &  & \emph{activate, excluding itself}// \\
16 & \textbf{end\_do} & \\
17 & $ilp(i) \leftarrow \frac{ilp(i)}{n-1}$ and $itp(i) \leftarrow \frac{itp(i)}{n-1}$ for all $i \in N$ & \\
18 & \textbf{return} $ilp(i)$ and $itp(i)$ for all $i \in N$ & \\
\hline
\end{tabular}
\caption{\label{Alg1} Algorithm for computing the ILP and ITP indices.}
\end{center}
\end{figure}

The algorithm starts initializing the $itp(i)$ variables, for all $i \in N$. Then, for each node $i \in N$, it generates a cascade of influences, trying to activate the nodes placed in the candidates list $L$. So, $i$ is the first node to enter the active nodes list $S$ and the activating candidates list $L$ is initialized with all the successors of $i$ in $G$. Then, while list $L$ is nonempty, we take from $L$ the node ($v$) with the lowest out-degree in $G$ and test it for activation. If $v$ passes the test, then we include it in $S$ and put in $L$ all the inactive successors of $v$. In addition, and if $v$ becomes active, we increase variable $itp(v)$ in one unit, meaning that one additional node ($i$) is able to activate $v$ through influence propagation. Then, node $v$ is removed from $L$, whether it entered $S$ or not. When $L$ becomes empty, the process terminates for the influential search started in node $i$, allowing to set the $ilp(i)$ result, which is equal to the number of nodes that were made active, excluding node $i$, divided by $n-1$.

Each new candidate to activate taken from list $L$ is selected according to the minimum out-degree of the nodes in $L$. The advantage of this option is to promote a fast activation of the easiest nodes. The execution times increase when priority is given to the nodes with largest out-degree.

\section{Computational tests and discussion} \label{Sec3:Compt}

In this section we use the ILP-ITP($\alpha$) algorithm for computing the ILP and ITP indices on a number of randomly generated and real-world graphs. We start describing the instances in the first subsection; then in Subsection \ref{Subsec3.2:Metrics}, we use a small known example to compare the ILP and ITP indices with other metrics usually considered in the literature. The computational results and comments on larger sized graphs are conducted in Subsection \ref{Subsec3.3:Res}.

All tests were run on an Intel Core i7-2600 with 3.40 GHz and 8 GB RAM. The experiments were performed under Microsoft Windows 10 operating system.
The algorithm described in Figure \ref{Alg1} was coded in Fortran and compiled on gfortran. Times are reported in seconds.

\subsection{Instances} \label{Subsec3.1:Inst}

In the tests conducted in this section, we consider two classes of instances: randomized and real-world. Randomized instances represent small-world networks, following the methodology described in \cite{WatStr1998}. Real-world instances are taken from known online repositories and used in a number of publications addressing virtual social networks. The repositories are: the Stanford Large Network Dataset Collection (SNAP, \cite{LesKre2014}), the Koblenz Network Collection (KONECT, \cite{Kun2017}), and from the Social Networks Security Research Group from the Ben-Gurion University of the Negev (BGU; \cite{LesCheRokElo2013}). All these instances are described next.

\medskip
Small-world randomly generated instances:

\medskip
The randomly generated instances are classified by nodes' average degree and sparsity. All instances involve $n=10,000$ nodes. Their initial average degrees are 10, 20 and 50. We follow the methodology described in \cite{WatStr1998} for generating small-world networks. Depending on the construction process, these graphs can have social network properties, as described in \cite{Bar2016} and \cite{GunRag2017}. The small-world construction of each network is initially made for the undirected version. Then, we use the undirected graph to build two additional oriented versions by substituting some of the edges by an arc in one of the two directions, while substituting the remaining edges by the two associated oriented arcs. The two oriented versions involve the substitution of $(2/3)m$ and $(1/3)m$ edges by a single directed arc, for $m$ the number of edges in the undirected counterpart. The edges selected to be oriented are taken at random. These networks are denoted by WS-\emph{k}-\emph{o}, with $k$ representing the initial average nodes' degrees ($k=10, 20 \mbox{ and } 50$) and $o$ indicating the proportion of bidirected links between pairs of nodes (edges in the original graph) that will remain, taking values $o=1.00, 0.66 \mbox{ and } 0.33$). In version $o=1.00$, all links are bidirected, representing the undirected graph; for $o=0.66$, the graph is oriented, keeping 66\% of the initial set of edges (as bidirected arcs) and 33\% of unidirected arcs; and for $o=0.33$ the graph is also oriented, keeping 33\% of the initial set of edges (as bidirected arcs) and 66\% unidirected arcs.

The Watts and Strogatz procedure starts with a regular graph with nodes' degree \emph{k}. Each edge of the graph is rewired, being reconnected, with probability \emph{p}, to another node chosen uniformly at random (duplicate edges are forbidden). The process is repeated for all original edges. Following a number of works in the literature, also addressing social networks (see, .e.g, \cite{GunRag2017,FisKahLeiMonRut2018,RagZha2019}), we consider the most destructive rewiring probability in the recommended range ($0.1 \leq p \leq 0.3$), setting $p=0.3$. This rewiring probability represents most closely the social networks studied by Watts and Strogatz (\cite{WatStr1998}).

\medskip
Table \ref{tab:t3} summarizes the main characteristics of the WS randomly generated graphs. The information concerning nodes' degrees and density is adapted to the correspondent version of the graph, considering the degrees' values for the undirected versions and the out-degrees ($g^+(i)$) for the oriented cases.

\smallskip {\small
\begin{table}[h]
  \centering
 {\footnotesize
 \begin{tabular}{l|ccccccc}
  \hline
   &  &  &  & \multicolumn{3}{c}{degree/out-degree} & \\
   & nodes ($n$) & edges/arcs ($m$) & density & min & average & max & type \\
  \hline
 WS-10-33 & 10,000 & 66,500 & 0.0007 & 0 & 6.65 & 15 & oriented \\
 WS-10-66 & 10,000 & 83,000 & 0.0008 & 2 & 8.30 & 16 & oriented \\
 WS-10-100 & 10,000 & 50,000 & 0.0010 & 5 & 10.00 & 18 & undirected \\
 WS-20-33 & 10,000 & 133,000 & 0.0013 & 3 & 13.30 & 25 & oriented \\
 WS-20-66 & 10,000 & 166,000 & 0.0017 & 6 & 16.60 & 28 & oriented \\
 WS-20-100 & 10,000 & 100,000 & 0.0020 & 12 & 20.00 & 30 & undirected \\
 WS-50-33 & 10,000 & 332,500 & 0.0033 & 17 & 33.25 & 50 & oriented \\
 WS-50-66 & 10,000 & 415,000 & 0.0042 & 20 & 41.50 & 59 & oriented \\
 WS-50-100 & 10,000 & 500,000 & 0.0100 & 38 & 50.00 & 66 & undirected \\
  \hline
\end{tabular}}
  \caption{{\small Main characteristics of the WS randomly generated graphs}} \label{tab:t3}
\end{table}}

\bigskip
Real-world instances:

\medskip
$\bullet$ Zachary karate club \cite{Zac1977}:

Data source: http://vlado.fmf.uni-lj.si/pub/networks/data/ucinet/ucidata.htm

This dataset was collected from the members of a university karate club by the sociologist Wayne Zachary in 1977 \cite{Zac1977}. Each node represents a member of the club and an edge between two members indicates that they are connected, generating an undirected graph. Each club member knew all the others, but the network only represents links between members who regularly interact outside the club. This network is widely used in a number of papers in the literature. Most of these works try to find the two groups of people into which the karate club split after an argument between the president (John A.) and an instructor (Mr. Hi). We use this network to compare the ILP and ITP indices with other known metrics in the literature.

\medskip
$\bullet$ Konect - Advogato network \cite{MasSalTom2009}:

Data source: http://konect.uni-koblenz.de/networks/advogato

The Advogato trust network is build using the Advogato online community platform for developers of free software. The original dataset has 3992 loops that were removed. The resulting graph is oriented, with 5,155 nodes (Advogato users) and 47,135 arcs (trust relationships, called a "certification" on Advogato).

\medskip
$\bullet$ Konect - Hamsterster network \cite{Kun2013}:

Data source: http://konect.uni-koblenz.de/networks/petster-friendships-hamster

This Network is based on friendships between users of the website hamsterster.com. The network is undirected, with 1,858 nodes (users) and 12,534 edges (friendships).

\medskip
$\bullet$ SNAP - ego-Facebook \cite{LesMca2012}:

Data source: https://snap.stanford.edu/data/index.html

This network represents social circles (circles of friends) from Facebook (anonymized), collected from survey participants using this online social network. Nodes are Facebook users and edges represent interactions between users. The dataset includes users' profiles, circles and ego networks. The resulting graph is undirected, with 4,039 nodes (users) and 88,234 edges (interactions). The anonymized process permits relating users by their affiliations but does not allow to identify those affiliations.

\medskip
$\bullet$ SNAP - email-EU-core network \cite{YinBenLesGle2007,LesKleFal2007}:

Data source: https://snap.stanford.edu/data/email-Eu-core.html

This network was generated using email data from a large European research institution. It used anonymized information about all incoming and outgoing email messages between members. The resulting graph is oriented, with 1,005 nodes (members of the research institution) and 25,571 arcs, where an arc $(i,j)$ in the graph indicates that member $i$ sent at least one email message to $j$, considering just email messages shared between members (the core).

\medskip
$\bullet$ SNAP - CollegeMsg temporal network \cite{PanOpsCar2009}:

Data source: https://snap.stanford.edu/data/CollegeMsg.html

This network involves an online social network at the University of California, Irvine. It is a temporal network based on private messages shared among members, derived from a dataset hosted by Tore Opsahl \cite{PanOpsCar2009}. Users could search the network for others and then initiate conversation based on profile information. The resulting graph is oriented, with 1,899 nodes (members) and 20,296 arcs, involving 59,835 messages shared along a given time frame. An arc $(i,j)$ means that user $i$ sent at least one message to $j$ within a given time frame.

\medskip
$\bullet$ BGU - Ning network \cite{LesCheRokElo2013}:

Data source: http://proj.ise.bgu.ac.il/sns/ning.html

Ning is a very large online community building platform for people and organizations to create social networks. This particular network is a snapshot of the friendship and group affiliation networks from Ning, harvested during September 2012. The resulting graph is oriented, with 10,298 nodes (members) and 76,262 arcs. It has 5,512 pairs of nodes linked by a single arc. The remaining arcs represent bidirected links among pairs of nodes. We have removed one loop from the original dataset.

\medskip
We have switched all arcs' orientation in the real-world networks because a link from node $i$ to node $j$ in the original dataset indicates that $i$ follows $j$, that is, $j$ is followed by $i$. Thus, according to the definition of the influence graph $G$ introduced in Section \ref{Sec1:Intro}, this link should be represented by arc $(j,i)$, that is, node $j$ influences node $i$.

Table \ref{tab:t4} summarizes the main characteristics of the graphs, using the same column labels considered in Table \ref{tab:t3}, besides "source", representing the online repository dataset.

\smallskip {\small
\begin{table}[h]
  \centering
 {\footnotesize
 \begin{tabular}{l|cccccccc}
  \hline
   &  &  &  &  & \multicolumn{3}{c}{degree/out-degree} & \\
   & source & nodes ($n$) & edges/arcs ($m$) & density & min & average & max & type \\
  \hline
 Zachary karate club & Konect & 34 & 78 & 0.1390 & 1 & 4.59 & 17 & undirected \\
 Advogato & Konect & 5,155 & 47,135 & 0.0018 & 0 & 9.14 & 721 & oriented \\
 Hamsterster & Konect & 1,858 & 12,534 & 0.0073 & 2 & 13.49 & 272 & undirected \\
 ego-Facebook & SNAP & 4,039 & 88,234 & 0.0108 & 1 & 43.69 & 1045 & undirected \\
 email-EU & SNAP & 1,005 & 25,571 & 0.0253 & 0 & 25.44 & 212 & oriented \\
 CollegeMsg & SNAP & 1,899 & 20,296 & 0.0056 & 0 & 10.69 & 137 & oriented \\
 Ning & BGU & 10,298 & 76,262 & 0.0007 & 0 & 7.41 & 633 & oriented \\
  \hline
\end{tabular}}
  \caption{{\small Main characteristics of the real world instances used in the computational tests}} \label{tab:t4}
\end{table}}

An important difference observed among the WS randomly generated instances and the real world examples here considered, is the range of variation of nodes' degrees. Although the average settings are comparable, the differences between minimum and maximum degrees in the real world examples are much larger, which is closer to the usual behavior of a social network.

\subsection{Other metrics} \label{Subsec3.2:Metrics}

The most usual metrics in the literature and in real-life problems belong to two classes: centralized and link topological metrics (see, e.g., \cite{KisBic2008, PenZhoCaoYuNiuJia2018}). Centralized metrics characterize the spread capabilities of the nodes and also describes nodes' proximity to the other players in the network; while link topological metrics emphasize important neighbors, benefiting from their relevancy. These metrics are also included in the measures list considered in Gephi \cite{BasHeyJac2009}. The lists in these classes are vast, particularly in the centralized group. In the present paper, we consider the selection described next.

Before, we must introduce the concept of \emph{path length} between two nodes, denoted by $p_{ij}$ for all pairs in $\left\{\{i,j\}: i,j \in N \mbox{ and } i \neq j \right\}$, where $p_{ij}$ is the length (number of arcs or edges) in the shortest path between $i$ and $j$ in $G$.

\bigskip
Centralized metrics:

\medskip
$\bullet$ degree centrality (\cite{Fou2012, PenZhoCaoYuNiuJia2018}):

The degree of a node $i \in N$ is the number of links incident on $i$. If the graph is oriented, we consider just its out-degree, as defined in Section \ref{Sec1:Intro}, that is, $g^+(i) = |\delta^+(i)|$, related with the strength of influence and the hurdle of a node, considered in (\ref{eqn:1}). We use the notation $g(i)$ when addressing the undirected case. Nodes with higher degree (more friends or more followers) are considered to be more influential.

\medskip
$\bullet$ eccentricity (\cite{HagHar1995, Fou2012}):

The eccentricity of a node $i \in N$ is the maximum path length between $i$ and any other node in the graph, that is, $ec(i) = \max_{j \in N}p_{ij}$.
The central nodes in the graph are those with the lowest eccentricity.

\medskip
$\bullet$ closeness centrality (\cite{OkaCheLi2008}):

The closeness centrality of node $i \in N$ is the inverse of the average path length between $i$ and all the other nodes in the graph, that is, $cc(i) = \frac{n-1}{\sum_{j \in N \setminus\{i\}} p_{ij}}$. Nodes with larger closeness centrality can reach sooner the other nodes in the graph, on average, being better positioned to spread the information.

\medskip
$\bullet$ betweenness centrality (\cite{Fre1977, Bra2001, BocLatMorChaHwa2006}):

The betweenness centrality (or load) of node $i \in N$ represents a measure of the number of shortest paths passing through node $i$, being defined by $bc(i) = \sum_{s,t \in N \setminus \{i\}, s \neq t} \frac{\sigma_{st}(i)}{\sigma_{st}}$, with $\sigma_{st}(i)$ representing the number of shortest paths between $s$ and $t$ passing through node $i$, and $\sigma_{st}$ representing the number of shortest paths between $s$ and $t$ in $G$. Nodes with larger betweenness centrality are in the "preferable" path between many pairs of nodes, acting as bridges. Under the assumption that information flows through shortest paths, these nodes are better placed to control most communications' traffic in the graph. In addition, the betweenness centrality value of node $i$ also indicates the disruption induced in the graph when $i$ is removed.

\bigskip
Link topological metrics (\cite{PenZhoCaoYuNiuJia2018}):

\medskip
$\bullet$ eigenvector centrality (\cite{Bon2007, Luetal2016}):

The eigenvector centrality of node $i \in N$ measures the level of influence of node $i$ based on the influential strength of their direct neighbors, being denoted by $ev(i)$. This means that the spreading ability of $i$ is enforced if it is connected to other influential nodes.
In this case, a person $i$ with few connections can have a high $ev(i)$ if those connections are also well-connected with others in the network. Nodes with higher eigenvector value are expected to be more influential in the network. The eigenvector metric is the same as degree centrality if all neighbor nodes have equal degree, namely on regular graphs. The eigenvector centralities can be computed iteratively through the expression $ev(i)=\frac{1}{\lambda} \sum_{j \in \delta^-(i)} ev(j)$, that is, $Ax=\lambda x$, where $A$ is the adjacency matrix, $x$ is the $ev$ vector and $\lambda$ is the largest eigenvalue in absolute value of $A$ ($\lambda \neq 0$), related to the principal eigenvector of $A$.

\medskip
$\bullet$ PageRank (\cite{BriPag2012, Luetal2016, Liuetal2017}):

The PageRank index, proposed by Sergey Brin and Lawrence Page \cite{BriPag1998, BriPag2012}, and used by Google's search engine, is a measure of the relevancy of a given web page in a webgraph (oriented, in general). It is also used in other network environments (\cite{Gle2015}), including social networks. As the eigenvector metric, this index is based on the number and relevancy of the hiperlinks a website receives. Thus, higher PageRank values correspond to more meaningful websites in a webgraph. The PageRank considers the number of inward arcs (i.e., converging links to a website) and the relevancy of the linkers (i.e., the PageRank of the adjacent converging websites). The PageRank index of a node $i$, here denoted by $pr(i)$ for all $i \in N$, is calculated using a recursive procedure, defined by $pr_t(i) = \beta \sum_{j \in \delta^-(i)} \frac{pr_{t-1}(j)}{g^+(j)} + (1-\beta)\frac{1}{n}$, where $\beta$ is the probability that a user clicks on a hyperlink on the current page; while $1-\beta$ is the probability of teleportation by typing the name of a web page of choice in the address bar, moving elsewhere. The procedure starts with $pr_1(i) = 1$ for all $i \in N$ and terminates when the PageRank values become stable. The final result is $pr(i) = pr_{tt}(i)$ for all $i \in N$, with $tt$ the total number of iterations of the recursive procedure. It has good convergence properties when $\beta$ is not too close to 1 (\cite{Gle2015}). The usually recommended value for $\beta$ is 0.85 (e.g., \cite{BriPag2012, Gle2015, Luetal2016}).

\medskip
$\bullet$ HITS (Hyperlink-Induced Topic Search) (\cite{Kle1999, Luetal2016}):

HITS was also developed to produce metrics on webgraphs, classifying websites into authoritative and hub pages, here denoted by $au(i)$ and $hu(i)$, respectively, for all $i \in N$. Authoritative pages are those with many incoming links from hub websites, exposing the value of the content of the page; while hub pages are those incident on many relevant authoritative pages. These two types of nodes are intimately related to each other, being calculated through a recursive procedure that should converge to stable authoritative and hub scores. The recursive expressions are: $au_t(i) = \sum_{\delta^-(i)} hu_{t-1}'(j)$ and $hu_t(i) = \sum_{\delta^+(i)} au_{t-1}'(j)$, with $au_t'(i) = \frac{au_t(i)}{\sqrt{\sum_{k \in N} \left( au_t(k)\right)^2}}$ and $hu_t'(i) = \frac{hu_t(i)}{\sqrt{\sum_{k \in N} \left( hu_t(k)\right)^2}}$ the normalized scores. The procedure starts with $au_1(i) = hu_1(i) = 1$ for all $i \in N$; and terminates when the scores of all nodes reach the steady state, controlled by a given threshold, denoted by $\varepsilon$ and set to 0.0001 in our computational tests. The final values are $au(i) = au_{tt}(i)$ and $hu(i) = hu_{tt}(i)$ for all $i \in N$, with $tt$ the total number of iterations of the recursive procedure. The two metrics' results are similar on undirected graphs. A good authoritative page should have many incoming links from many good hubs, that is, being linked from pages known as hubs for information; and a good hub page should point to many good authoritative pages, that is, linked to pages that are considered to be authorities on the subject. The nodes with higher authoritative value are the most central in the graph.

\bigskip
Unlike other authors, we chose to include the eigenvector metric within the link topological class because it is build upon neighborhood influence, as PageRank and HITS.

All these centralized measures are focused on direct indistinct neighbors or on the path length distance to the other nodes in the graph. They do not reveal information spread based on nodes' influence strength over their neighbors. This influence strength over the neighbors is essential to model influence spread cascades, used to show each node's influence capability over the entire network, performed by the ILP measure here proposed.

The same way, the selected link topological metrics described above are based on direct neighborhood influence, distinguishing the relevancy of the various neighbors. These neighbors also benefit from their neighbors influence and so on, producing a kind of influence propagation towards the nodes. However, in this cases, the relevancy of a member is entirely build upon the relevancy of its neighbors, ignoring its own ability to decide on a message propagation scheme. In effect, message propagation capability depends on the strength of inward neighbors, but it then must break the node's own hurdle to further propagate the message. This point establishes the main difference between the link topological metrics described above and influence propagation considered on Threshold and Cascade models, including the ILP and ITP indices architecture.
Instead of just reading neighbors' direct influence, the ILP index uses the network topology and the message viral power to truly simulate message propagation. On a different perspective, the ITP index describes each node capability as a consumer of information flowing in the graph. It may resemble the betweenness centrality measure, but instead of receiving the flow due to its geodesic position in the graph, it receives the flow by true influence propagation cascades started on each node in the graph, being also sensitive to the message viral power.

To detach these differences, we use the Zachary karate club network described in Subsection \ref{Subsec3.2:Metrics}, represented by an undirected graph.

Table \ref{tab:t10} in the Appendix shows the ILP and ITP indices' values for all the nodes in the Zachary's club graph, for $\alpha = 1.5 \mbox{ and } 3.0$. It also includes all the results produced by the metrics described above, calculated using Gephi \cite{BasHeyJac2009}.

\begin{figure}[h]
\begin{center}
\begin{tabular}{ccc}
\includegraphics[scale=0.25]{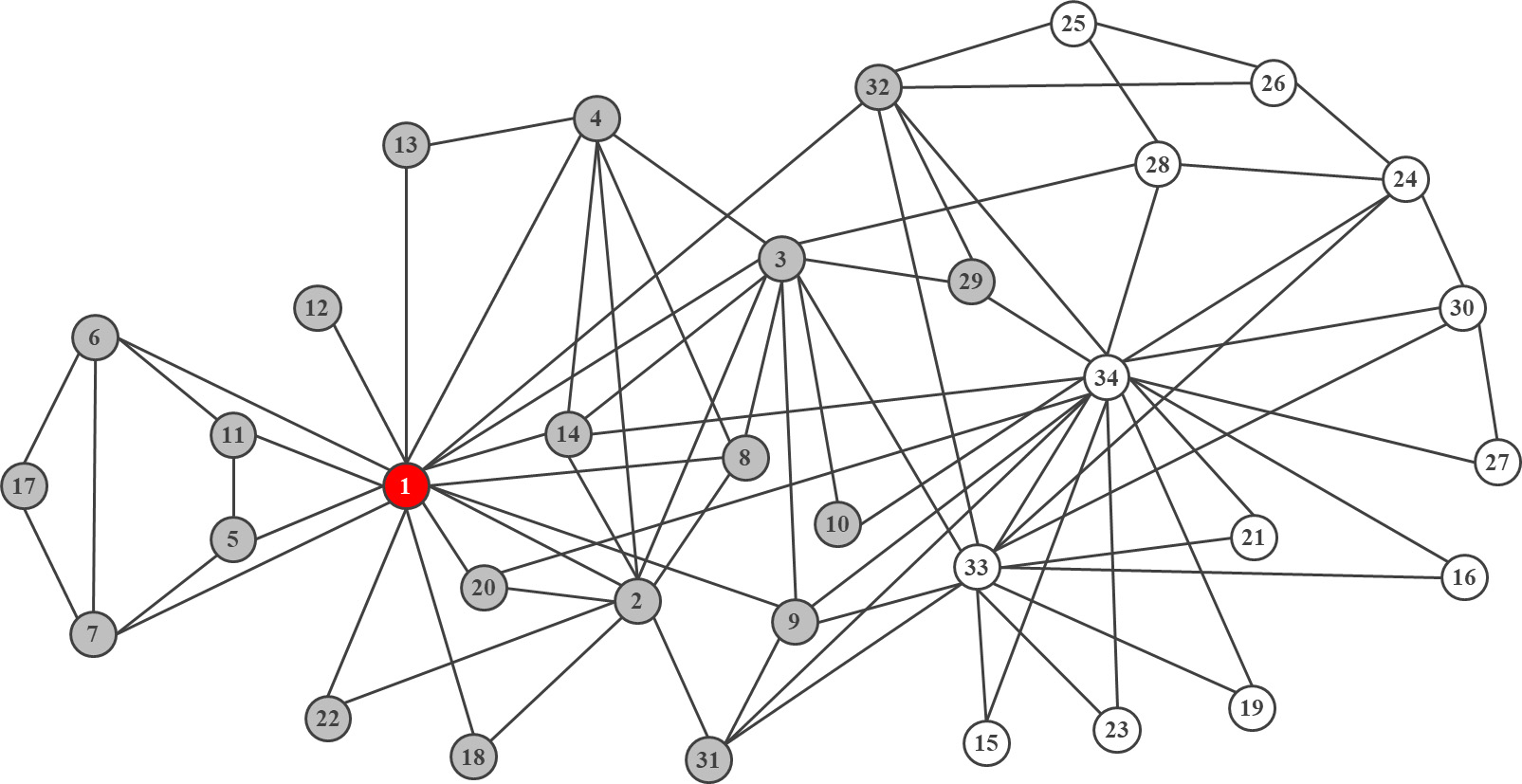} & \mbox{\hspace*{0,1cm}} & \includegraphics[scale=0.25]{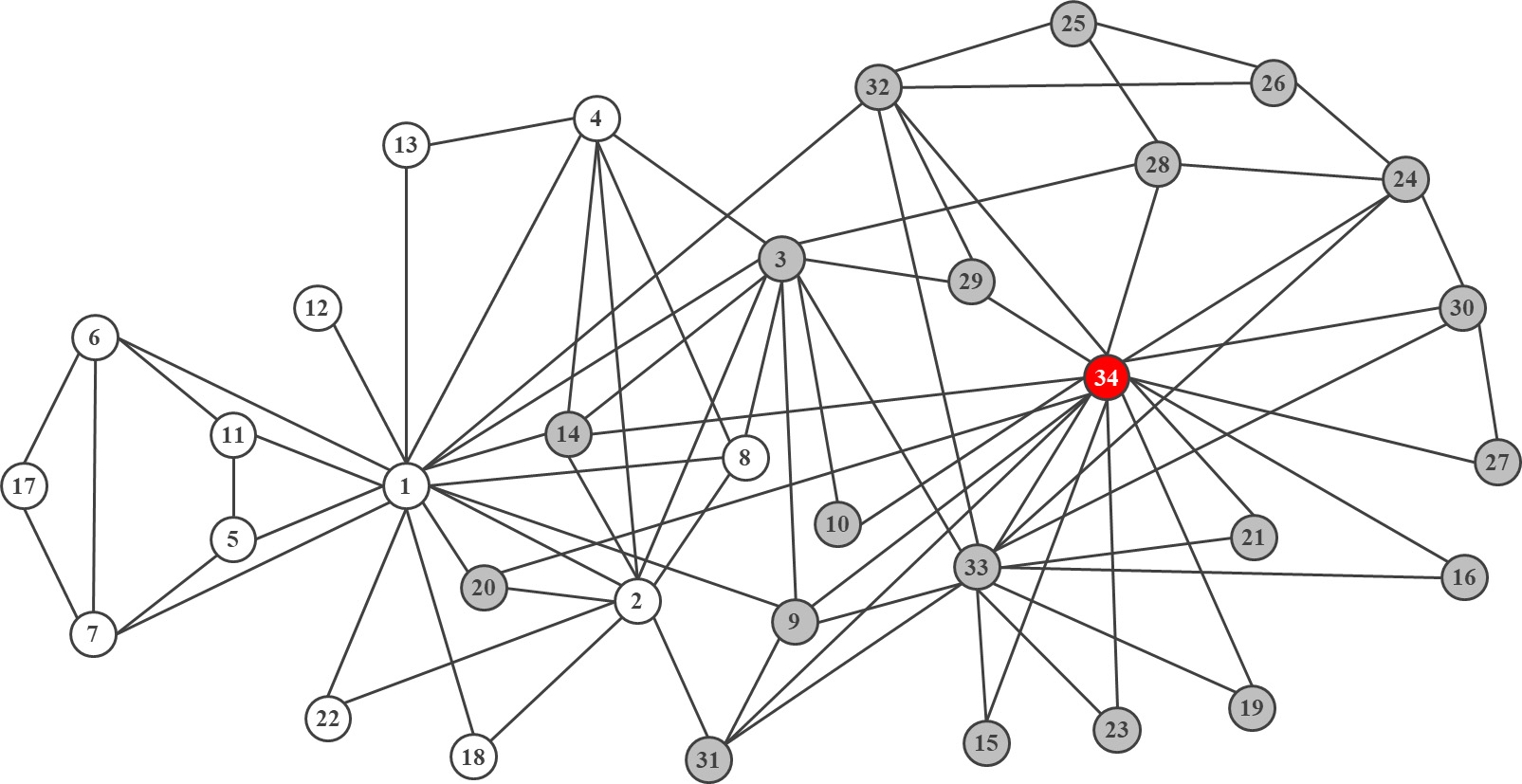} \\
(a) & & (b) \\
\end{tabular}
\caption{\label{fig2} ILP index solution with $\alpha = 3.0$, for node 1 (in (a)) and node 34 (in (b)) as the origins, for the Karate club Zachary's graph.}
\end{center}
\end{figure}

Starting with the ILP index results, when $\alpha = 1.5$, representing a rather viral message, for instance, an opinion about an adversary, there are 5 members able to spread out this opinion throughout the entire network (members: 1 (Mr. Hi), 2, 3, 33 and 34 (John A.)). The other members have low or null influence capability in the graph. This is a typical behavior of this index that separates the members in two groups: strong and low/null message launchers. However, when $\alpha = 3.0$, representing a less viral message, for instance, a non-consensual opinion about one of the leaders (Mr. Hi or John A.), the message propagation should be harder because they are all members of the same club. In this case, only node 1 (Mr. Hi) and node 34 (John A.) can spread the message, being able to cover only 20 other members, when starting in each one of them, represented in Figure \ref{fig2} (a) and (b), for Mr. Hi and John A., respectively. There is another member (node 33) that can also do some damage, being able to reach 9 other members, but clearly with lower strength compared to the two leaders.

The two solutions in Figure \ref{fig2} also show that there are some members being reached by the two potential influencers: Mr. Hi and John A., namely members 3, 9, 10, 14, 20, 29, 31 and 32.
Some of these members stayed with Mr. Hi (3, 9, 14, 20) and the others stayed with John A. (10, 29, 31, 32) after the club's fission.

Concerning the ITP index results, when the message is volatile ($\alpha = 1.5$) there are two nodes (17 and 26) that are more influenceable than the others, being reached by 7 members of the club. These two members have great potential for being consumers or collectors of information in the network, which may suggest that they should be observed closely for all sort of message propagation in the graph, namely gossips or fake news. However, for messages with less spread capability ($\alpha = 3.0$), namely classified information, the members to observe are no longer the former, but 9, 10, 20, 29 and 31. Curiously, they all belong to the subset of those influenced by both sides, as observed above.

Compared to the other metrics, the ILP and ITP indices incorporate the capability to classify the message being spread. In effect, the viral strength of the message is a key issue for its spreadability, while the previously mentioned metrics make no distinction on this matter. If we observe the degree information of each node, it provides close answers for the more volatile message (with $\alpha = 1.5$), but it helps less when the message has more friction (with $\alpha = 3.0$), namely on nodes 2, 3 and 33. However, nodes' degree can complement the ILP index information, revealing, for instance, that some lower degree members (2 and 3) are also able to cover the entire network when the message is volatile. These members can be easier to convince to act as influencers in a marketing campaign, for instance. In addition, among the members with the lowest eccentricity there are a number of them with low or null ability to spread the message, namely members 4, 9, 14, 20 and 32. Instead, John A. (member 34) has higher eccentricity. Also, closeness centrality detaches members 9, 14, and 32 with low average path length (below $\frac{1}{0.52}$) and very low or null capability to be message spreaders. On the other hand, member 17 has high average path length ($\frac{1}{0.28}$) while being a good consumer/collector of volatile messages when $\alpha = 1.5$, with $ipt(17) = 0.21$. Then, the betweenness centrality metric detaches members 3 and 32 as interesting bridges for the flow passing among nodes in the graph, but they have low capability to be good influencers, specially for less viral messages ($\alpha = 3.0$). This is curious because they can be highly disruptive if removed from the graph. In addition, the stronger ability of the members as consumers/collectors (index ITP) is not followed by higher values of betweenness centrality. Now, considering the link topological metrics, they all bring similar results for this example, detaching members 1 and 34, and also members 2, 3 and 33 for their relevancy in the graph, which is directly related with the number of neighbors and their weight. However, they lack indicating how far this relevancy can go considering message propagation. In effect, they also stress nodes 4, 9, 14 and 32 (as the eccentricity and closeness centrality metrics) with moderate relevancy, although these members have low or null ability for message spreading in the graph. To conclude, these observations stress that the new indices should not be used to substitute the known metrics, but used in conjunction with those metrics to complement the decision process on influencers selection in social networks.

In the next subsection we provide a few more observations involving the metrics here described using a larger sized graph.

\subsection{Computational results} \label{Subsec3.3:Res}

This section describes the computational tests on the ILP-ITP($\alpha$) algorithm described in Section \ref{Sec2:Alg} for generating the ILP (Individual Launching Power) and ITP (Individual Target Potential) indices on two given classes of instances. These classes involve the randomized and real-world instances selected in Subsection \ref{Subsec3.1:Inst}.

\medskip
Table \ref{tab:t6} reports the ILP and ITP values on the WS randomly generated graphs, involving $n=10,000$ nodes. The table includes the execution times (in seconds) of the algorithm and the percentage of nodes that are able to cover (activate) at least 99\% of the remaining nodes in the graph, representing strong candidates to be influencers, using the ILP index results. The tests were conducted considering the following values for the hurdle coefficient: $\alpha =$ 1.0, 1.5 and 2.0, indicating the virality level of a message generated in each of the nodes.

\smallskip {\small
\begin{table}[h]
  \centering
 {\footnotesize
 \begin{tabular}{lcc|ccc|ccc}
  \hline
   &  &  & \multicolumn{3}{c|}{percentage of strong influencers} & \multicolumn{3}{c}{execution times (sec.)} \\
 instance  & density & type & $\alpha = 1.0$ &  $\alpha = 1.5$ & $\alpha = 2.0$ &  $\alpha = 1.0$ & $\alpha = 1.5$ & $\alpha = 2.0$ \\
  \hline
 WS-10-33 & 0.0007 & oriented & 80.99 & 4.23 & 0.00 & 1853 & 70 & $<1$ \\
 WS-10-66 & 0.0008 & oriented & 86.57 & 14.57 & 0.35 & 2443 & 371 & 7 \\
 WS-10-100 & 0.0010 & undirected & 90.31 & 16.87 & 0.44 & 3472 & 552 & 13 \\
 WS-20-33 & 0.0013 & oriented & 91.70 & 21.01 & 5.32 & 4227 & 1730 & 231 \\
 WS-20-66 & 0.0017 & oriented & 94.23 & 23.05 & 1.66 & 4923 & 1214 & 89 \\
 WS-20-100 & 0.0020 & undirected & 94.87 & 6.77 & 0.03 & 7183 & 502 & 3 \\
 WS-50-33 & 0.0033 & oriented & 97.18 & 21.01 & 0.37 & 9808 & 2142 & 38 \\
 WS-50-66 & 0.0042 & oriented & 98.24 & 8.53 & 0.09 & 11023 & 993 & 11 \\
 WS-50-100 & 0.0100 & undirected & 97.87 & 0.14 & 0.00 & 16792 & 24 & $<1$ \\
  \hline
\end{tabular}}
  \caption{{\small Percentage of strong influencers and execution times of the ILP-ITP($\alpha$) algorithm on the WS randomly generated graphs, considering $\alpha =$ 1.0, 1.5 and 2.0.}} \label{tab:t6}
\end{table}}

As expected, all the ILP index results observed using the WS graphs divide the nodes in two classes: launchers and non-launchers. The same way, the results with index ITP also found two classes of nodes: targets and non-targets. The breaking point that separates launchers from non-launchers represents the minimum proportion of nodes that launchers can cover. This value was observed to be higher than 0.9959 in all experiments reported in Table \ref{tab:t6}, except for instances WS-10-33 and WS-50-100 for $\alpha = 2.0$.

Considering the results in this table, when the message is very viral ($\alpha=1.0$), most nodes are able to act as strong launchers in the graph, especially when it becomes denser, as expected. As an example, suppose a social network (connected) of football (soccer) supporters. If we think of a very viral message in this network, for instance, "Ronaldo returns to Manchester", it can easily reach most nodes of the network if launched by any member, no matter its strength, in most cases. A similar result holds for a less viral message ($\alpha = 1.5$) when $k=10$, but it suddenly changes when the graph becomes denser. In effect, for $k=20$ the number of strong launchers is lower in the undirected graph (WS-20-100) compared with the lower density instances (WS-20-33 and WS-20-66); and this behavior is even clearer for the $k=50$ instances, where the number of stronger launchers decreases along with the density increase of the graphs, which is somehow unexpected. This performance is even more noticeable for the less viral message cases, with $\alpha = 2.0$. Actually, although density growth increases the number of connections, it also makes the nodes stronger in their own hurdle, turning message dissemination harder to pass. These small-world networks are particularly sensitive to this aspect, due to the homogeneity of their nodes' degrees. As observed in the forthcoming tests, this low variation on nodes' degrees is not typical in a social network, which may cast doubt on the suitability of small-world artificial graphs to simulate social networks.

Now, considering the larger sized real-world instances proposed in Subsection \ref{Subsec3.1:Inst}, we show in Tables \ref{tab:t8} and \ref{tab:t9} the results of the ILP and ITP indices returned by the ILP-ITP($\alpha$) algorithm, considering the following values for the hurdle coefficient: $\alpha =$ 1.0, 1.5, 2.0 and 3.0. The algorithm was run for the entire graphs, despite the fact that most of them are not connected (or strong connected in the oriented cases). The sizes of these connected (or strongly connected) components (in percentage of nodes over the entire graph) and the execution times of the algorithm for the various hurdle coefficients are reported in Table \ref{tab:t7}.

\smallskip {\small
\begin{table}[h]
  \centering
 {\footnotesize
 \begin{tabular}{lc|ccccc}
  \hline
   &  &  & \multicolumn{4}{c}{execution times (in seconds)} \\
 instance & type & largest component (in \%) &  $\alpha = 1.0$ & $\alpha = 1.5$ & $\alpha = 2.0$ & $\alpha = 3.0$ \\
  \hline
 Advogato & oriented & 60.91 & 12 & 6 & 4 & 2 \\
 Hamsterster & undirected & 96.23 & 3 & 2 & 1 & $<1$ \\
 ego-Facebook & undirected & 100.00 & 189 & 125 & 88 & 30 \\
 email-EU & oriented & 79.90 & 2 & 1 & $<1$ & $<1$ \\
 CollegeMsg & oriented & 68.14 & 2 & 1 & 1 & $<1$ \\
 Ning & oriented & 87.37 & 69 & 32 & 18 & 7 \\
  \hline
\end{tabular}}
  \caption{{\small Sizes (in percentage) of the largest connected (strongly connected) components in the real-world instances; and the execution times of the ILP-ITP($\alpha$) algorithm for $\alpha =$ 1.0, 1.5, 2.0 and 3.0.}} \label{tab:t7}
\end{table}}

The execution time of the algorithm is influenced by the size and density of the graph, but also by the hurdle coefficient. In effect, when the hurdle coefficient increases, the number of launcher nodes diminishes, being reflected on a lower execution effort by the algorithm. In addition, the largest connected (or strongly connected) component has almost the initial graph size on the undirected instances, being smaller on the oriented counterparts, especially on the Advogato graph. Despite these differences, we chose to run and report the tests on the original graphs because it is closer to reality.

Here again, all the ILP index results on the RW graphs divided the nodes in two classes: launchers and non-launchers. The same way, the results with index ITP found two classes of nodes: targets and non-targets. We consider again the breaking point that separates launchers from non-launchers as the minimum proportion that launcher members can cover, here denoted as \emph{minimum influential breaking point} (mibp in short). We also consider the breaking point that separates targeting from non-targeting nodes as the minimum proportion of nodes covering target nodes, denoted by \emph{minimum targeting breaking point} (mtbp in short). Thus, Table \ref{tab:t8} reports the percentage of launcher nodes and the mibp values found in each instance and for each hurdle coefficient. Table \ref{tab:t9} represents the percentage of targeting nodes and the mtbp values for the same instances and the same hurdle coefficients.

\smallskip {\small
\begin{table}[h]
  \centering
 {\footnotesize
 \begin{tabular}{l|cccc|cccc}
  \hline
   & \multicolumn{4}{c|}{percentage of launchers} & \multicolumn{4}{c}{min influential breaking point (mibp)} \\
 instance  & $\alpha = 1.0$ &  $\alpha = 1.5$ & $\alpha = 2.0$ & $\alpha = 3.0$ &  $\alpha = 1.0$ & $\alpha = 1.5$ & $\alpha = 2.0$ & $\alpha = 3.0$ \\
  \hline
 Advogato & 21.51 & 11.70 & 7.29 & 3.16 & 0.7241 & 0.7115 & 0.6995 & 0.6814 \\
 Hamsterster & 32.88 & 20.61 & 12.70 & 5.71 & 0.9553 & 0.9467 & 0.9435 & 0.9413 \\
 ego-Facebook & 73.31 & 52.34 & 36.37 & 15.23 & 1.0000 & 0.9861 & 0.9861 & 0.9854 \\
 email-EU & 52.64 & 36.42 & 26.27 & 12.14 & 0.8137 & 0.8137 & 0.8118 & 0.8078 \\
 CollegeMsg & 25.43 & 12.06 & 6.79 & 2.05 & 0.6965 & 0.6907 & 0.6886 & 0.6797 \\
 Ning & 14.58 & 7.44 & 4.47 & 1.93 & 0.8757 & 0.8642 & 0.8527 & 0.8196 \\
  \hline
\end{tabular}}
  \caption{{\small ILP index results on the real-world selected instances, considering $\alpha =$ 1.0, 1.5, 2.0 and 3.0.}} \label{tab:t8}
\end{table}}

The mibp percentage on these instances is lower, in general, compared to the WS instances results, especially on the Advogato and CollegeMsg datasets, probably due to the smaller size of the largest strongly connected components. On the other instances, the launcher nodes can cover more than 80\% of the nodes.

If observing the largest oriented graph (Ning), when the message is very viral ($\alpha=1.0$), 14.58\% of the nodes (1501 members) can be classified as launchers and the nodes in this group can reach at least 87.57\% (9018 nodes) of the entire set of members. The remaining 85.42\% (non-launcher nodes) can reach no more than 83 other nodes in the graph. However, if the message becomes less viral, with $\alpha=1.5$, the successful launcher nodes falls to 7.44\% (766 members) and each of these nodes can cover 86.42\% (8900 nodes) of the graph, or more. The remaining 9532 non-launcher nodes can reach less than 184 other members in the graph. Further, if the message has low virality (with $\alpha=3.0$), the percentage of launcher nodes is 1.93\%, representing only 199 members that are able to cover at least 81.96\% (8440 nodes) of the graph. The remaining 98.07\% members (nonlaunchers) can only cover at most 509 other nodes.
An additional observation considering, for instance, the $\alpha = 1.0$ case, the mibp of the launchers (0.8757, representing 9018 nodes) is bigger than the largest strongly connected component in that graph (87.37\%, that is, 8997 nodes). The reason for this is that the message can propagate across nodes in and out of the largest strongly connected component, namely among nodes in weakly connected components.

Considering a different case, if we observe the largest undirected graph (ego-Facebook network), which is entirely connected, it has 73.31\% launcher members that are able to cover the entire graph through influence propagation if the message is very viral, with $\alpha=1.0$. The remaining 26.69\% members can reach no more than 60 other nodes in the graph, thus, being non-launchers. Once again, if the virality of the message decreases, considering $\alpha=1.5$, the percentage of launcher nodes decreases to 51.34\%, each of which being able to cover 98.61\% of the graph; and if the message becomes harder to pass, with $\alpha=2.0$, the percentage of launcher nodes decreases further, to 36.37\%, falling even deeper (15.23\%) if the virality of the message is further decreased. These 15.23\% launchers are 615 Facebook members that are able to reach (individually) 98.54\% of the other nodes in the graph, at least.

\begin{figure}[h]
\begin{center}
\includegraphics[scale=0.465]{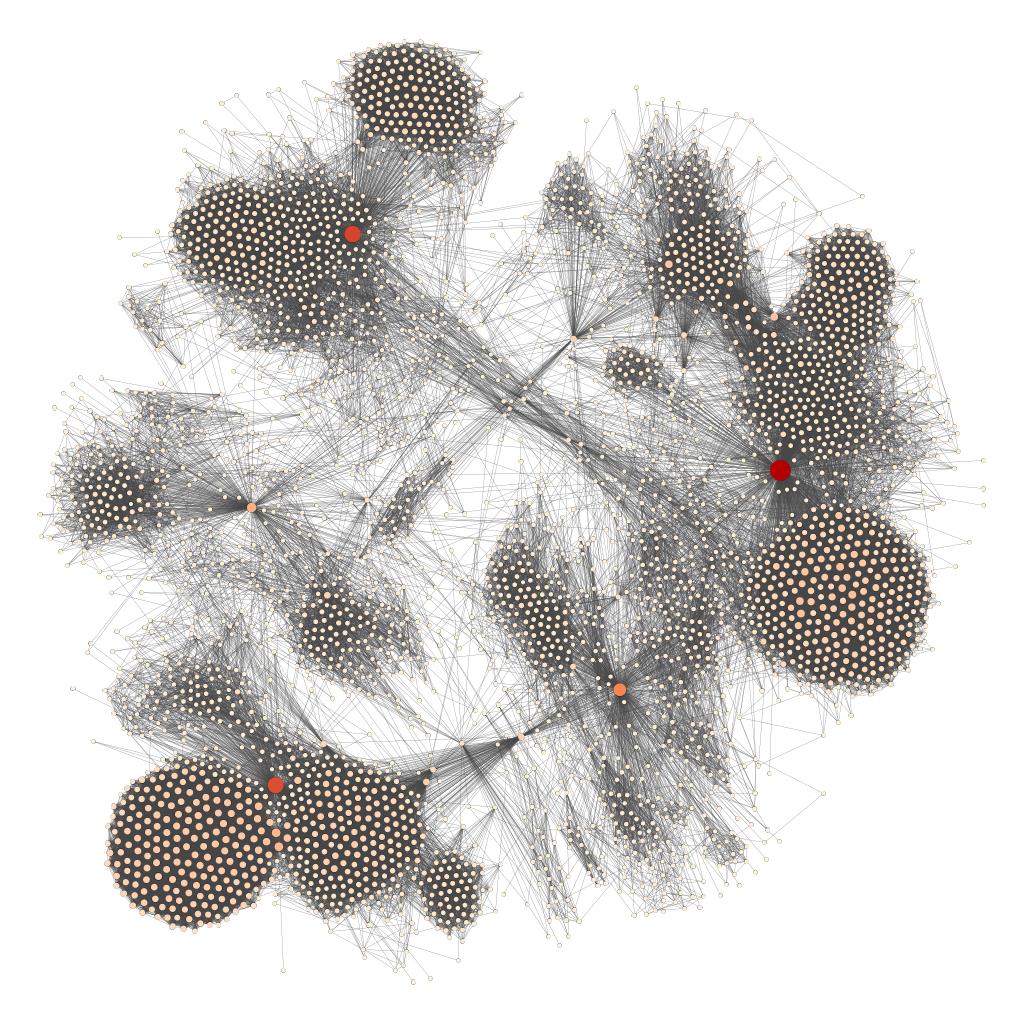}
\caption{\label{fig3} ego-Facebook network, build using Gephi.}
\end{center}
\end{figure}

The detached launcher members able to cover almost the entire graph are still too much if we are intended to choose some of them to propagate a message or initiate a marketing campaign. Therefore, we propose complementing the information with some of the metrics discussed in Subsection \ref{Subsec3.2:Metrics}. To assist on this discussion we show in Figure \ref{fig3} an image with the ego-Facebook instance, where the nodes' color (from red to light yellow) and size are proportional to their degree in the graph. The network was build using Gephi \cite{BasHeyJac2009}. Observing this image, a natural choice for strong influencers (as launchers) would be those with larger degree, according to the criteria used in the construction of the activating condition (\ref{eqn:1}) described in Section \ref{Sec1:Intro}. In fact, the five nodes with largest degree are 108, 1685, 1913, 3438 and 1, with degrees 1045, 792, 755, 547 and 347, respectively. These 5 nodes are also on top of the list for closeness centrality, betweenness centrality and PageRank, indicating that they are central on communication and neighbors influence. These are probably the most relevant players in the graph, but they should also be the more expensive if we think about a financial incentive to pay these members to support a marketing campaign or to decide sending a message. In effect, they all belong to the launchers' list determined by the ILP index. However, and still thinking about the cost to pay to these members, are there strong launcher nodes that may cost less? A possible answer to this question can arise from the lower degree nodes, or other metric, still belonging to the launchers' list. In effect, among the nodes in this list, we can find a number of members with degree below 30 (nodes 679, 3081, 3232 and 991) which may represent good candidates to act as launchers (influencers) in practice. Probably due to their position in the graph, these apparently weak members are so effective on influential spread as node 108 that exhibits the strongest degree ($g(108)=1045$) in the graph. Note that the average degree in this graph is 43.69 and the standard-deviation is 52.41. A curious aspect to mention involves node 568 that has the lowest eccentricity ($ec(568)=4$), node degree slightly above average ($g(568)=63$) and very high betweenness centrality (above 750,000), suggesting that it could be placed in a privileged position as an influencer. Yet, it is classified in the non-launchers class by the ILP index for virality level $\alpha=3.0$, as it is able to reach only 21 other nodes in the graph. So, the other known metrics may not be tailored for assessing influence propagation on their own, but their performance can be improved if used together with the ILP index information.

Besides instances ego-Facebook and email-EU, the launchers' groups on all other networks are relatively small. The number of members in these groups should fall below 1\% when $\alpha > 3.0$ in most of the studied graphs. This percentage is close to the number of nodes that are able to launch an effective influence cascade over the network, detached in \cite{GoeAndHofWat2016} and based on a very extensive variety of contents launched on Twitter. As shown in our experiments, these proportions are strongly influenced by the viral power of the message ($\alpha$) and the network topological nature.

\smallskip {\small
\begin{table}[h]
  \centering
 {\footnotesize
 \begin{tabular}{l|cccc|cccc}
  \hline
   & \multicolumn{4}{c|}{percentage of targets} & \multicolumn{4}{c}{min targeting breaking point (mtbp)} \\
 instance  & $\alpha = 1.0$ &  $\alpha = 1.5$ & $\alpha = 2.0$ & $\alpha = 3.0$ &  $\alpha = 1.0$ & $\alpha = 1.5$ & $\alpha = 2.0$ & $\alpha = 3.0$ \\
  \hline
 Advogato & 72.42 & 71.15 & 69.95 & 68.15 & 0.2150 & 0.1168 & 0.0728 & 0.0314 \\
 Hamsterster & 95.53 & 94.67 & 94.35 & 94.13 & 0.3285 & 0.2057 & 0.1265 & 0.0565 \\
 ego-Facebook & 100.00 & 98.61 & 98.61 & 98.54 & 0.7330 & 0.5233 & 0.3777 & 0.1521 \\
 email-EU & 81.39 & 81.39 & 81.19 & 80.80 & 0.5259 & 0.3635 & 0.2620 & 0.1205 \\
 CollegeMsg & 69.67 & 69.09 & 68.88 & 67.98 & 0.2540 & 0.1201 & 0.0674 & 0.0200 \\
 Ning & 87.57 & 86.42 & 85.27 & 81.96 & 0.1457 & 0.0743 & 0.0446 & 0.0192 \\
  \hline
\end{tabular}}
  \caption{{\small ITP index results on the real-world selected instances, considering $\alpha =$ 1.0, 1.5, 2.0 and 3.0.}} \label{tab:t9}
\end{table}}

Now, concerning the ITP index results reported in Table \ref{tab:t9}, most nodes in the graphs belong to the targets group, with a slight exception on instances CollegeMsg and Advogato, possibly due to their lower connectivity properties. However, most of these targeting nodes are reachable by a relatively small number of members, except on the ego-Facebook and email-EU graphs. For instance, the targeting group in the ego-Facebook graph includes all members when the message is very viral ($\alpha=1.0$) and these members are reachable by at least 73.30\% (2961 members) other nodes in the graph. In this case, there are no non-targeting members. However, when the message has low virality, with $\alpha=3.0$, the targeting group includes 98.54\% members (3980 nodes), but these members are only reachable by 15.21\% other nodes in the graph. The remaining 1.46\% non-targeting members (59 nodes) can be reached by no more than 0.07\% of the nodes, that is, at most 3 members. The targeting class, in particular, includes a large variety of members, all of them acting as message consumers (or collectors). This group of members may possibly concentrate the usual targets of fake news that can be used to start the tracking process of fake news' origins. To further explore the selection within the targeting members' group, we may use the betweenness centrality metric as an additional filter. If we focus again on the $\alpha=3.0$ virality level and still on the ego-Facebook network, the nodes with largest betweenness centrality value (larger than 1 million) are, again, members 108, 1685, 3438, 1913, 1086 and 1, among those in the targeting nodes' class. Curiously, all these nodes are both good launchers and good targets. Also, they are among top degree members, except node 1086 ($g(1086)=66$), so with no novelty, they are strong players in the graph. However, they are also very heavy players that can belong to the expensive nodes' class.

To conclude, an interesting and expected observation in these tests shows that the percentage of launchers is almost the same as the mtbp proportion; and the percentage of targets is also close to the mibp proportion. This illustrates that the nodes able to reach the target members are basically the launchers; and the nodes that are reached by the launchers are mostly the targets.

\section{Conclusions} \label{Sec4:Conclusions}

In the present paper, we have considered an entirely deterministic process for characterizing adoption and influence using an activating condition based on the Linear Threshold (LT) model.
The activating condition uses the topological information of the graph, namely the nodes' out-degree (or degree if it is undirected) and a single parameter - the hurdle coefficient - for classifying the viral propagation strength of the message/product under consideration. Thus, it does not depend on personal information of the users and hence can be applied easily in practice.

Based on this process, we have proposed an algorithm that produces two influence propagation indices for online social networks: the ILP (Individual Launching Power) and the ITP (Individual Target Potential). The ILP provides a clear division of the nodes into launchers and non-launchers, with very low distinction inside each group. Each of the launchers can cover most of the graph through influential cascades, reaching more than 70\% of the members in most social networks used in our tests. The size of the launchers' group diminishes significantly with the hurdle coefficient increase, reflecting the natural virality variation of the message or marketing campaign (message virality weakens with the hurdle coefficient increase). It also depends on the social network topology. When the message has low virality (with $\alpha > 3.0$), the launchers group's size is possibly lower than 1\%, being in line with \cite{GoeAndHofWat2016} that observed that less than 1\% of the influential cascades are able to pass beyond the very next neighbors.

This partition of the nodes into launchers and non-launchers can be used as a first filter for the selection of the best candidates for influence propagation. Then, we can use other metrics in the literature to choose the members with the best characteristics.
In effect, if we choose influencers on a social network without this filter, considering, for instance, just the nodes' degree information, we would possibly tend to select the members with largest degree (individuals with more followers). However, those are also typically the more expensive if used in a marketing campaign or to pass a message. As we have observed in the computational tests here conducted, there are other members in the launchers' group that are able to reach a similar performance as the mentioned very strong members, but having a significantly lower degree (having much less followers). These lower degree influencers are probably much less expensive, being able to perform almost as well as the highest degree members on the given network. So, instead of searching for the largest degree nodes, we recommend looking for the lower degree ones but belonging to the launchers' group, earlier found by the ILP index. These are probably the "ordinary influencers" (individuals who exert average or even less-than-average influence) observed in \cite{BakHofMasWat2011} when studying influence propagation on Twitter. This process can be conducted using other centralized or link topological metrics besides nodes' degree, depending on the kind of members we are looking for.

We have also observed that the ILP index information obtained from artificial small-world networks (instances WS, generated according to \cite{WatStr1998}) produce lower sized launchers' groups when the density of the graph increases, especially when the hurdle coefficient is higher (lower virality messages). Although higher density graphs have more connections, offering more chances for message propagation; it also makes the nodes stronger in their own hurdle, which turns them harder to collaborate on message transmission. This last observation may justify the mentioned unexpected behavior found on the small-world graphs, which is possibly justified by the low variation of their nodes' degrees. This low variation is not typical on social networks, which may cast doubt on the suitability of small-world artificial graphs to simulate social networks.

The ITP index, instead, divides the nodes into targeting and non-targeting members. According to our tests, most nodes in the graphs belong to the targets group (more than 70\% on most social networks). However, each of those targeting nodes are targeted by a small number of members (no more than 25\% of the nodes, in most cases), falling sharply with the hurdle coefficient increase, that is, when the message becomes less viral. These targeting nodes can represent compulsive consumers of information in online networks, which may include fake news' easy targets. As mentioned above, the targeting group is very large, so, once again, we can use it as a first filter and then resort to other available metrics to assist on the search for specific members selection profiles.

On another perspective, the ILP and ITP indices could be used to restrict the set of candidate seed nodes in most Influence Maximization problems based on the Linear Threshold model. That restricted subset should focus on strong influencers (launchers) with low chance of sharing common nodes in their propagation cascades.

An additional aspect to stress involves the choice of adequate values for the hurdle parameter $\alpha$. The present work introduces a brief discussion on this matter, but further work is needed, specially on real-practice environments for adequately tuning this parameter. In the meanwhile, we recommend considering sensitivity analysis on $\alpha$ in any new real-world instance.

In future works, the ILP and ITP indices can be discussed using the Independent Cascade model \cite{KemKleTar2003} instead of the Linear Threshold model here considered. Also, and in this line, instead of assuming that all individuals are equally receptive for a message/product to be launched, we could consider these assumptions to be ruled stochastically.

\section*{Acknowledgements}
Pedro Martins acknowledges support from the Portuguese National Funding: Fundação para a Ciência e a Tecnologia - FCT, under the project UIDB/04561/2020.

\section*{Appendix}

Table \ref{tab:t10} shows the ILP and ITP indices' values for the Zachary's club graph. It also includes the results of the metrics described in Subsection \ref{Subsec3.2:Metrics}, calculated using Gephi \cite{BasHeyJac2009}.

\smallskip {\small
\begin{table}[h]
  \centering
 {\footnotesize
 \begin{tabular}{c|cc|cc|cccc|cccc}
  \hline
  nodes & \multicolumn{2}{|c}{$ilp(i)$} & \multicolumn{2}{|c}{$itp(i)$} & \multicolumn{4}{|c}{centralized} & \multicolumn{4}{|c}{link topological} \\
  $i$ & $\alpha = 1.5$ & $\alpha = 3.0$ & $\alpha = 1.5$ & $\alpha = 3.0$ & $g(i)$ & $ec(i)$ & $cc(i)$ & $bc(i)$ & $ev(i)$ & $pr(i)$ & $au(i)$ & $hu(i)$ \\
  \hline
1 & 1 & 0.61 & 0.12 & 0 & 16 & 3 & 0.57 & 231.07 & 0.96 & 0.1 & 0.36 & 0.36 \\
2 & 1 & 0.09 & 0.12 & 0.03 & 9 & 3 & 0.49 & 28.48 & 0.7 & 0.05 & 0.27 & 0.27 \\
3 & 1 & 0.06 & 0.12 & 0.06 & 10 & 3 & 0.56 & 75.85 & 0.84 & 0.06 & 0.32 & 0.32 \\
4 & 0.06 & 0.03 & 0.15 & 0.03 & 6 & 3 & 0.46 & 6.29 & 0.56 & 0.04 & 0.21 & 0.21 \\
5 & 0 & 0 & 0.15 & 0.03 & 3 & 4 & 0.38 & 0.33 & 0.21 & 0.02 & 0.08 & 0.08 \\
6 & 0.06 & 0 & 0.18 & 0.03 & 4 & 4 & 0.38 & 15.83 & 0.23 & 0.03 & 0.08 & 0.08 \\
7 & 0.06 & 0 & 0.18 & 0.03 & 4 & 4 & 0.38 & 15.83 & 0.23 & 0.03 & 0.08 & 0.08 \\
8 & 0 & 0 & 0.18 & 0.03 & 4 & 4 & 0.44 & 0 & 0.45 & 0.02 & 0.17 & 0.17 \\
9 & 0 & 0 & 0.15 & 0.09 & 5 & 3 & 0.52 & 29.53 & 0.61 & 0.03 & 0.23 & 0.23 \\
10 & 0 & 0 & 0.15 & 0.09 & 2 & 4 & 0.43 & 0.45 & 0.27 & 0.01 & 0.1 & 0.1 \\
11 & 0 & 0 & 0.15 & 0.03 & 3 & 4 & 0.38 & 0.33 & 0.21 & 0.02 & 0.08 & 0.08 \\
12 & 0 & 0 & 0.15 & 0.03 & 1 & 4 & 0.37 & 0 & 0.14 & 0.01 & 0.05 & 0.05 \\
13 & 0 & 0 & 0.18 & 0.06 & 2 & 4 & 0.37 & 0 & 0.22 & 0.01 & 0.08 & 0.08 \\
14 & 0 & 0 & 0.15 & 0.06 & 5 & 3 & 0.52 & 24.22 & 0.6 & 0.03 & 0.23 & 0.23 \\
15 & 0 & 0 & 0.15 & 0.06 & 2 & 5 & 0.37 & 0 & 0.27 & 0.01 & 0.1 & 0.1 \\
16 & 0 & 0 & 0.15 & 0.06 & 2 & 5 & 0.37 & 0 & 0.27 & 0.01 & 0.1 & 0.1 \\
17 & 0 & 0 & 0.21 & 0.03 & 2 & 5 & 0.28 & 0 & 0.07 & 0.02 & 0.02 & 0.02 \\
18 & 0 & 0 & 0.15 & 0.06 & 2 & 4 & 0.38 & 0 & 0.25 & 0.01 & 0.09 & 0.09 \\
19 & 0 & 0 & 0.15 & 0.06 & 2 & 5 & 0.37 & 0 & 0.27 & 0.01 & 0.1 & 0.1 \\
20 & 0 & 0 & 0.15 & 0.09 & 3 & 3 & 0.5 & 17.15 & 0.4 & 0.02 & 0.15 & 0.15 \\
21 & 0 & 0 & 0.15 & 0.06 & 2 & 5 & 0.37 & 0 & 0.27 & 0.01 & 0.1 & 0.1 \\
22 & 0 & 0 & 0.15 & 0.06 & 2 & 4 & 0.38 & 0 & 0.25 & 0.01 & 0.09 & 0.09 \\
23 & 0 & 0 & 0.15 & 0.06 & 2 & 5 & 0.37 & 0 & 0.27 & 0.01 & 0.1 & 0.1 \\
24 & 0.03 & 0 & 0.15 & 0.06 & 5 & 5 & 0.39 & 9.3 & 0.41 & 0.03 & 0.15 & 0.15 \\
25 & 0 & 0 & 0.18 & 0.03 & 3 & 4 & 0.38 & 1.17 & 0.16 & 0.02 & 0.06 & 0.06 \\
26 & 0 & 0 & 0.21 & 0.03 & 3 & 4 & 0.38 & 2.03 & 0.17 & 0.02 & 0.06 & 0.06 \\
27 & 0 & 0 & 0.18 & 0.03 & 2 & 5 & 0.36 & 0 & 0.2 & 0.02 & 0.08 & 0.08 \\
28 & 0 & 0 & 0.15 & 0.03 & 4 & 4 & 0.46 & 11.79 & 0.36 & 0.03 & 0.13 & 0.13 \\
29 & 0 & 0 & 0.18 & 0.09 & 3 & 4 & 0.45 & 0.95 & 0.35 & 0.02 & 0.13 & 0.13 \\
30 & 0.03 & 0 & 0.15 & 0.06 & 4 & 5 & 0.38 & 1.54 & 0.36 & 0.03 & 0.13 & 0.13 \\
31 & 0 & 0 & 0.15 & 0.09 & 4 & 4 & 0.46 & 7.61 & 0.46 & 0.02 & 0.17 & 0.17 \\
32 & 0.09 & 0 & 0.15 & 0.06 & 6 & 3 & 0.54 & 73.01 & 0.52 & 0.04 & 0.19 & 0.19 \\
33 & 1 & 0.27 & 0.12 & 0.03 & 12 & 4 & 0.52 & 76.69 & 0.83 & 0.07 & 0.31 & 0.31 \\
34 & 1 & 0.61 & 0.12 & 0 & 17 & 4 & 0.55 & 160.55 & 1 & 0.1 & 0.37 & 0.37 \\
  \hline
\end{tabular}}
  \caption{{\small ILP and ITP indices and the results of the metrics described in Subsection \ref{Subsec3.2:Metrics} for the Zachary's club graph}} \label{tab:t10}
\end{table}}

\end{document}